\documentclass[aps,prb,twocolumn,showpacs,superscriptaddress,amsmath,amssymb,citeautoscript]{revtex4}
\usepackage{graphicx}
\usepackage{dcolumn}
\usepackage{multirow}
\usepackage{bm}
\usepackage{bbold}
\usepackage{hyperref}
\usepackage[capitalize]{cleveref}
\usepackage{amsmath}
\usepackage{commath}
\usepackage{nicefrac}
\usepackage{float}
\usepackage{verbatim}\usepackage{color}
\usepackage[FIGTOPCAP]{subfigure}

\begin{document}

\title{Fingerprints of Orbital Physics in Magnetic Resonant Inelastic X-ray Scattering}

\author{Pasquale Marra}
\address{Institute for Theoretical Solid State Physics, IFW Dresden, Helmholtzstrasse 20, 01069 Dresden, Germany}

\begin{abstract}
Orbital degrees of freedom play a major role in the physics of many strongly correlated transition metal compounds. However, they are still very difficult to access experimentally, in particular by neutron scattering.
We propose here how to reveal \emph{orbital} occupancies of the system ground state by \emph{magnetic} resonant inelastic x-ray scattering (RIXS).
This is possible because, unlike in neutron scattering, the intensity of the magnetic excitations in RIXS depends essentially on the symmetry of the orbitals where the spins are in.
\end{abstract}

\maketitle

\section{Introduction}
\quad

Fermi gases and Fermi liquids can be well described in terms of momentum phase space. In the latter systems, electrons can interact even strongly but, as long as one is able to integrate out interactions, they can be described as a Fermi gas of independent (non interacting) quasiparticles. 
However, this is not always the case. In Mott insulators and other strongly correlated systems local degrees of freedom have to be taken into account. 
In fact, in the regime of charge commensuration and strong interactions, electron states are almost localized, and the charge degree of freedom is frozen in the low energy limit. But, due to correlations, electron spin and angular momentum excitations can still propagate into the system, and therefore non trivial collective behaviors can emerge.

However, unlike spin, orbital angular momentum in correlated materials turns out to be very subtle. While examples of magnetic order are known since millennia (magnetite), correlated orbital states have attracted attention only recently~\cite{Tokura2000}. 
On the other hand, a direct probe of orbital degrees of freedom seems to be challenging. Due to its quenching in almost all relevant $3d$ transition metal oxides~\cite{VanVleck1932}, orbital angular momentum does not contribute to magnetic moment. Macroscopically, that implies that, unlike magnetization, the order parameter of a correlated orbital system is difficult to observe. For the same reason, orbital angular momentum is hardly detectable even on a microscopic scale. Neutron scattering, for instance, is almost blind to orbital degrees of freedom, since neutrons can interact only via magnetic coupling~\cite{Kim2011}. 

Having in mind these general considerations, the relevance of the Resonant Inelastic X-ray Spectroscopy (RIXS)~\cite{Groot1998, Ament2009, Haverkort2010, Ament2011} in studying orbital degrees of freedom in transition metal oxides can be better understood. In fact, due to the electromagnetic coupling, photons interact strongly with local orbital angular momenta and, moreover, the inelastic nature of this technique allows one to measure directly orbital excited states, e.g., $dd$ excitations~\cite{Ulrich2008, Forte2008b}.  Direct observations of orbital excitations however are often controversial because typically such $dd$ excitations couple strongly to the lattice,  so that their energy dispersion cannot be unequivocally identified.

However, RIXS spectroscopy allows one also to probe \emph{indirectly} orbital degrees of freedom, by looking at magnetic excitations. At a transition metal ion $L_2$ or $L_3$ edge, the core hole in the intermediate state lies deep in the $2p$ shell, where the spin orbit coupling is strong enough to allow spin flip transitions in the final state. Since orbital angular momentum and spin are strongly coupled, spin flip transitions cross sections depend strongly on the local orbital state. In this way, by measuring the magnetic excitations one can shed light on the orbital nature of the ground state of a strongly correlated system. More precisely, the polarization dependence of RIXS intensities of spin flip transitions allows one to determine the symmetries of the occupied orbital states.

\section{Magnetic RIXS at ${\rm Cu}^{2+}$ $L_{2,3}$ edges}
\quad

\begin{figure}%
	\centering
	\includegraphics[trim= 0mm 0mm -10mm 0mm, clip, width=1\columnwidth]{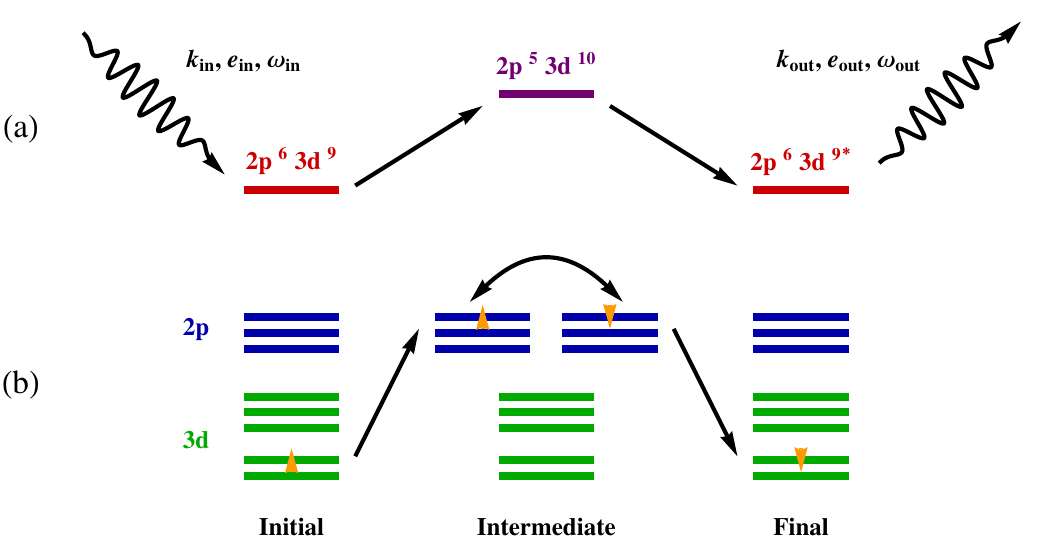}
	\label{fig:DirectRIXS}
	\caption{
A simple sketch of the direct RIXS process at the Cu$^{2+}$ $L_{2,3}$ edge. In (a) the initial, intermediate and final states are shown, each one specified by its electronic configuration, while in (b) these state configurations are represented in the \emph{hole} picture, in the specific case of a pure spin flip transition. 
}
\end{figure}

In direct RIXS spectroscopy (cf. Fig.~\ref{fig:DirectRIXS}), e.g., at a transition metal $L_{2,3}$-edge, an incoming photon raises a core electron ($2p$) into an empty valence state ($3d$). 
Eventually, an electron from a valence state decays to fill the core hole by the emission of an outgoing photon. The final state may coincide with the initial one, although there is a finite probability that the system is left in an excited state, e.g., with different orbital occupancies ($dd$ excitation) or, due to the spin orbit coupling, with a different spin (magnetic excitation). By measuring the incoming and outgoing x-ray energies and momenta, we obtain the energy $\omega=\omega_{\rm out} - \omega_{\rm in}$ and the momentum ${\bf k}={\bf k}_{\rm out} -{\bf k}_{\rm in}$ of the excitation left into the system.

\begin{table}
$
\begin{array}{l|l}
	3z^2-r^2	& 	-\frac16 e^{\imath \phi} \left[ \sin{\theta}(e_{xy}-e_{yx})+ \right.\\[1mm]
			&\left.	+2 (\cos{\phi}\cos{\theta}-\imath\sin{\phi})(e_{yz}-e_{zy})+\right.\\[1mm]
			&\left.	+2(\sin{\phi}\cos{\theta}+\imath\cos{\phi})(e_{zx}-e_{xz})\right]	\\[1mm]
	x^2-y^2	/ xy	& 	\frac12 e^{\imath \phi}\sin{\theta}(e_{xy}-e_{yx})		\\[1mm]
	yz		& 	-\frac12 e^{\imath \phi}(\cos{\phi} \cos{\theta} - \imath \sin{\phi})(e_{yz}-e_{zy})	\\[1mm]
	zx		& 	-\frac12 e^{\imath \phi}(\sin{\phi}\cos{\theta} + \imath \cos{\phi})(e_{zx}-e_{xz})
\end{array}
$
	\caption{
Scattering amplitudes $\vert O_{{\bf j}, {\bf e}}^{\downarrow\uparrow} \vert^2$ for pure spin flip transitions of a single Cu$^{2+}$ ion at $L_{2,3}$ edge for a quantization axis with inclination $\theta$ and azimuth $\phi$ and for arbitrary polarization $e_{\alpha\beta}=e^{in}_{\alpha}e^{out*}_{\beta}$.
}
	\label{tab:SpinFlip}
\end{table}

The general form of the magnetic RIXS cross section at a transition metal $L_{2,3}$ edge is~\cite{Haverkort2010, Ament2011}
\begin{equation}\label{eq:crosssection}
	 I_{\bf e} ( {\bf k}, \omega ) = \!\!
	\lim_{\delta \rightarrow 0^+}\! \mathrm{Im} 
	\langle 0 \vert  
	\hat{O}^\dag_{{\bf k}, {\bf e}} 
	\frac{1}{\hbar\omega + E_0 - H + \imath \delta}  
	\hat{O}_{{\bf k}, {\bf e}} 
	\vert  0 \rangle 
	,
\end{equation}
where $\mathbf{e}=\mathbf{e}^{\rm in}\cdot(\mathbf{e}^{\rm out})^{\dagger}$ is the tensor that describes the incoming and outgoing photon polarization, $H$ is the Hamiltonian describing 3$d$ valence electrons with ground state $\vert 0 \rangle $ and energy $E_0$,  and $\hat{O}_{{\bf k}, {\bf e}} = 1/\sqrt{N}\sum_{\bf j} \hat{O}_{{\bf j}, {\bf e}} \exp(\imath {\bf k \cdot j})$ is the Fourier transformed RIXS transition operator.
Within the dipole approximation the RIXS operator can be expressed as~\cite{Ament2011}
\begin{equation}\label{RIXSOperator}
	\hat{O}_{{\bf j}, {\bf e}} = 
	\sum_{\alpha \beta} {e}_{\alpha \beta} \hat{D}^\dag_{{\beta}, {\bf j}} \left(\frac{1}{\hbar\omega_{res}+E_0 - H_{c} +\imath \Gamma}\right) \hat{D}_{{\alpha}, {\bf j}} 
	,
\end{equation}
where $\hbar\omega_{res}$ is the incoming photon energy tuned to a specific absorption edge, $\Gamma$ the core hole lifetime, $\hat{D}_{\alpha {\bf j}}$ the components of the dipole operator~\cite{Ament2011} (greek indexes span over $x$, $y$ and $z$ hereafter), $H_{c}$ the intermediate state Hamiltonian, and where the operator sandwiched in parenthesis is the intermediate state propagator $\hat{G}_{\bf j}$.

In general, the intermediate state Hamiltonian can be very complicated, even considering only local interactions, since electrons in the valence states are interacting with themselves and with the potential created by the core hole via Coulomb repulsion, and the independent electron approximation is no longer realistic. 
\begin{figure}[t]
	\includegraphics[width=1\columnwidth]{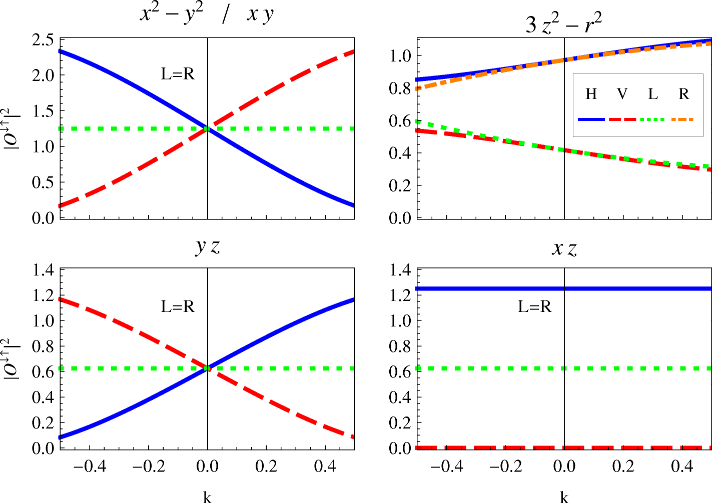} 
	\caption{
Amplitudes $\vert O_{{\bf j}, {\bf e}}^{\downarrow\uparrow} \vert^2$ for pure spin flip transitions for a single Cu$^{2+}$ ion at $L_3$ edge, with spin in the $xy$ plane, and for several orbital ground state symmetries. The scattering plane is (100), with a scattering angle 90$^\circ$. Amplitudes depend on the polarization and are plotted against the momentum $k=k_x=k_y$ transferred on the $xy$ plane. Results for distinct incoming polarizations are shown [circular left (L) and right (R) polarization and linear horizontal (H) and vertical (V) polarization], while outgoing ones are averaged out.
}
	\label{fig:SpinFlip}
\end{figure}
Direct RIXS at Cu$^{2+}$ $L_{2,3}$ edges however turns out to be easier, since the entire RIXS process can be seen in a very simple way as a single particle problem in the hole picture (see Fig.~\ref{fig:DirectRIXS}). The initial configuration of a Cu$^{2+}$ ion is $2p^6 3d^9$, i.e., one hole in the $3d$ shell [cf. Fig.~\ref{fig:DirectRIXS}(a)]. The incoming photon raises an electron from the core shell to the valence one, so that the electronic configuration is now $2p^5 3d^{10}$, with one hole in the core shell and with the valence shell fully occupied. In the final state, one photon is emitted and the electronic configuration comes back to the initial one. In the hole picture [see Fig.~\ref{fig:DirectRIXS}(b)], the incoming photon excites the hole from the $3d$ to the $2p$ shell. This excited state eventually decays, and the hole in the core shell comes back to the $3d$ valence shell by emitting a photon. If one neglects long range interactions between the core hole and the surrounding ions, the only relevant interaction in the intermediate state is the spin orbit one. The energy levels of the core hole intermediate state are therefore split into two subsets, with total angular momentum ${\bf J}=1/2$ and ${\bf J}=3/2$, that corresponds to the resonant edges $L_2$ and $L_3$. 

Following~\cite{Ament2011}, the core hole propagator can be written in terms of these core hole states. For large spin orbit coupling, only core hole states which are resonantly excited contribute to the propagator, while the others become negligible, so that 
\begin{equation}
	\hat{G}_{\bf j}=
	\sum_n\frac{ \vert n \rangle \langle n \vert }{\hbar\omega_{res}+E_0 - E_n +\imath \Gamma}
	\simeq
	\frac{1}{\imath \Gamma}\!\sum_{E_n=\hbar \omega_{res}}\!\!\!\! \vert n \rangle \langle n \vert ,
\end{equation}
where $\vert n \rangle$ is the core hole state with energy $E_n$.
Moreover, due to spin orbit coupling, these states are a superposition of states with different spin, so that the propagator can be written as~\cite{Haverkort2010}
$\hat{G}_{\bf j}=\left(\frac12+\hat{S}_{\bf j}^z\right) \hat{G}_{\bf j}^{\uparrow\uparrow} + \left(\frac12-\hat{S}_{\bf j}^z\right) \hat{G}_{\bf j}^{\downarrow\downarrow} + \hat{S}_{\bf j}^+ \hat{G}_{\bf j}^{\uparrow\downarrow} + \hat{S}_{\bf j}^- \hat{G}_{\bf j}^{\downarrow\uparrow}$,
where the projected propagators $\hat{G}_{\bf j}^{\sigma\sigma'}$ contain only operators acting on the orbital part of the core hole state wavefunction.
Introducing the operator $\hat{M}_{\gamma \bf j}=\sum_{\alpha\beta}\epsilon_{\alpha\beta\gamma} p^{\dagger}_{ \alpha,{\bf j}} p^{}_{\beta,{\bf j}}$, where $\epsilon_{\alpha\beta\gamma}$ is the Levi-Civita symbol and $p^{\dagger}_{ \alpha,{\bf j}}$ the creation operator of the 2$p$ core hole in the $p_{\alpha}$ orbital state, and writing the ladder spin operators in terms of the spin components, one obtains:
\begin{equation}
	\hat{G}_{\bf j} = 
	-\imath c_1 + c_2 \ \hat{\bf S}_{\bf j} \cdot \hat{\mathbf M}_{\bf j}
 	,
\end{equation}
where $c_1$ and $c_2$ are constants depending on the resonant edge. 
For pure spin flip transitions, the RIXS operator in Eq.~(\ref{RIXSOperator}) is then:
\begin{equation}\label{RIXSOperatorFinal}
	\hat{O}_{{\bf j}, {\bf e}} = \sum_{\alpha \beta} {e}_{\alpha \beta} 
	\hat{D}^\dag_{{\beta}, {\bf j}} \left(\hat{\bf S}_{\bf j} \cdot \hat{\mathbf M}_{\bf j} \right) \hat{D}_{{\alpha}, {\bf j}}
	.
\end{equation}
The full spin and rotational symmetry of the RIXS process is thus uncovered, and one can easily evaluate cross sections of spin excitations for any orientation of the initial valence state spin and for different orbital ground states, as shown in Tab.~\ref{tab:SpinFlip}. Due to spin orbit coupling, spin flip cross sections depend strongly on the orbital ground state (cf. Tab.~\ref{tab:SpinFlip}). This is particularly clear if one analyzes the polarization dependence of single Cu$^{2+}$ ion spin flip transition amplitudes $\vert O_{{\bf j}, {\bf e}}^{\downarrow\uparrow}(d) \vert^2=\vert \langle \downarrow d| O_{{\bf j}, {\bf e}} |\uparrow d\rangle \vert^2$, for different orbital ground states $d$, shown in Fig.~\ref{fig:SpinFlip}. 

The difference between the elongated ($3z^2-r^2$) and the planar orbitals ($x^2-y^2$ and $t_{2g}$) stands out in Tab.~\ref{tab:SpinFlip} and Fig.~\ref{fig:SpinFlip}. While in the latter case, circular left and right polarizations are equivalent, in the former case a non vanishing circular dichroism appears (cf. Fig.~\ref{fig:SpinFlip}). 
This is due to the fact that, for such an elongated orbital ground state, different spin flip transitions channels are allowed. As a consequence, for any choice of polarization and spin direction, spin flip transitions are always possible and, moreover, those different channels  can interfere in a non trivial way as soon as one considers circular polarization.

\section{Conclusions}
\quad 
We have shown that by measuring the scattering intensities of \emph{magnetic} spin flip excitations, Resonant Inelastic X-ray Scattering (RIXS) allows one to probe \emph{orbital} degrees of freedom in transition metal compounds and, more precisely, to determine the orbital occupancies of the ground state. This is possible because in the direct RIXS spin and orbital degrees of freedom are strongly coupled in the core hole state so that, unlike in inelastic neutron scattering, the magnetic scattering intensities strongly depend on the symmetry of the orbitals that spins occupy. 

\begin{acknowledgments}
I thank  Krzysztof Wohlfeld, Jeroen van den Brink, and V. Bisogni for fruitful discussions.
\end{acknowledgments}

\end{document}